\documentclass[usenatbib]{mnras}
\usepackage[T1]{fontenc}
\usepackage{ae,aecompl}
\usepackage{upgreek}
\usepackage{graphicx}	
\graphicspath{{images/}}
\usepackage{amsmath}	
\usepackage{amssymb}	
\usepackage{aas_macros}

\title[Secular evolution of VW~LMi]{Secular changes in the orbits of the quadruple system VW~LMi}
\author[T. Pribulla et al.]{T.~Pribulla,$^{1,2,3}$ E.~Puha,$^4$ T. Borkovits,$^{5,6}$ J. Budaj,$^1$ Z. Garai,$^{1,2,3}$ E. Guenther, $^7$ \newauthor L. Hamb\'alek,$^1$ R. Kom\v{z}\'{\i}k,$^1$ E.~Kundra,$^1$ Gy.~M. Szab\'o,$^{2,3}$ M. Va\v{n}ko$^1$\\
$^{1}$Astronomical Institute of the Slovak Academy of Sciences, 059 60 Tatransk\'a Lomnica, Slovakia, E-mail: pribulla@ta3.sk\\
$^2$ELTE Gothard Astrophysical Observatory, 9700 Szombathely, Szent Imre h. u. 112, Hungary\\
$^3$MTA-ELTE Exoplanet Research Group, 9700 Szombathely, Szent Imre h. u. 112, Hungary\\
$^{4}$Dept. of Astronomy, Physics of the Earth, and Meteorology, FMPH, Comenius University, Mlynsk\'a dolina,
842~48 Bratislava, Slovakia\\
$^5$Baja Astronomical Observatory of Szeged University, 6500 Baja, Szegedi u., Kt. 766, Hungary\\
$^6$Konkoly Observatory, Research Centre for Astronomy and Earth Sciences, H-1121 Budapest, Konkoly Thege Mikl\'os \'ut 15-17, Hungary\\
$^7$Th\"{u}ringer Landessternwarte, Sternwarte 5, 077~78 Tautenburg, Germany
}

\date{Accepted XXX. Received YYY; in original form ZZZ}
\pubyear{2019}

\begin{document}
\label{firstpage}
\pagerange{\pageref{firstpage}--\pageref{lastpage}}
\maketitle

\begin{abstract}
VW~LMi is the tightest known quadruple system with 2+2 hierarchy. It consists of a W~UMa-type eclipsing binary ($P_{12}$ = 0.47755 days) and another detached non-eclipsing binary ($P_{34}$ = 7.93 days) orbiting around a common center of mass in about $P_{1234}$ = 355 days. We present new observations of the system extending the time baseline to study long-term perturbations in the system and to improve orbital elements. The multi-dataset modeling of the system (4 radial-velocity curves for the components and the timing data) clearly showed an apsidal motion in the non-eclipsing binary at a rate of 4.6 degrees/yr, but no other perturbations. This is consistent with the nearly co-planarity of the outer, 355-day orbit, and the 7.93-day orbit of the non-eclipsing binary. Extensive N-body simulations enabled us to constrain the mutual inclination of the non-eclipsing binary and the outer orbits to $j_{34-1234} < 10$ degrees.     
\end{abstract}

\section{INTRODUCTION}
\label{intro}

VW~Leo~Minoris is a bright ($V$ = 8.06, $\alpha_{2000}$ = 11 02 51.9, $\delta_{2000}$ = +30 24 54) hierarchical quadruple system composed of F and G spectral-type stars (see Table~\ref{tab1}). Its photometric variability with a period of 0.477547 days and amplitude of about 0.41 mag was found by the Hipparcos mission \citep{esa}. It was classified as a W UMa-type eclipsing binary. First ground-based photometric measurements of VW~LMi were performed by \cite{dumitrescu1}. Later analysis of these light curves (LCs) and determination of system's parameters \citep{dumitrescu2} proved that VW~LMi is a contact binary. VW~LMi was included in the Fourier analysis of Hipparcos LCs of close eclipsing binaries of \citet{selam}, who confirmed it as a contact binary and found approximate geometrical elements as: fill-out factor $f$ = 0.40, mass ratio $q$ = 0.25 and the inclination angle $i$ = 72.5$\degr$. The resulting parameters are incorrect because of the large and at that time unknown third light. New photometric elements of VW~LMi were also obtained by \citet{sanchez} who re-analysed their older $BV$ observations \citep{gomez} taking into account the multiplicity of the system. The most recent photometric analysis of \citet{djurasevic}, based on new $BVR_{\rm c}$ observations, indicated a dark spot on the secondary component. Assuming spectroscopic mass ratio the authors found inclination angle $i_{12}$ = 78.1$\pm0.3\degr$, fill-out factor 0.504 and passband-dependent third light as $l_3/(l_1+l_2+l_3)$ = 0.24 ($B$ passband), 0.26 ($V$ passband) and 0.28 ($R$ passband) which indicates that the detached pair is of a later spectral type than the contact binary. They determined the distance to the system as 130$\pm$2 pc, which corresponds to $\pi$ = 7.69$\pm$0.12 mas.

Long-term spectroscopy (1998-2005) of VW~LMi by \citet{pribulla2006} at the David Dunlap Observatory (DDO) conclusively showed that the contact eclipsing binary (hereafter CEB) with a period of $P_{12}$ = 0.47755 days is gravitationally bound to a detached non-elisping binary (hereafter DNEB) with a period of $P_{34}$ = 7.93 days. The binaries revolve in a tight mutual outer orbit with $P_{1234}$ = 355 days. Later, \citet{pribulla2008} determined the masses of all four components by simultaneous modelling of 4 radial-velocity (RV) curves and timing variability of the CEB caused by the light-time effect (LiTE) in the outer mutual orbit. No further spectroscopic observations have been published. Although the system is rather bright, and the expected maximum separation of the binaries is about 10~mas, it has not been resolved interferometrically yet. 

Although  no astrometric acceleration of VW~LMi was detected by the Hipparcos, the re-analysis of the Hipparcos and the Gaia data by \citet{brandt} clearly shows astrometric acceleration especially in the declination. While the Gaia proper motion is $\mu_\delta = -6.08\pm0.24$ mas/yr, the position difference of VW~LMi as observed by Hipparcos and Gaia results in $\mu_\delta = -4.753\pm0.024$ mas/yr, i.e., 1.3 mas/yr difference significant at more than 5$\sigma$ level.

In this paper we present and analyze new photometric and spectroscopic observations. Our goals are: (i) to perform simultaneous modelling of available minima times of the CEB and RVs of all four components, (ii) to determine absolute parameters of all four components, (iii) to perform numerical integration of the system in order to study its long-term orbital evolution, (iv) to search for secular perturbations of the inner orbits (those of the subsystem binaries), and to (v) estimate the mutual inclinations of the outer and inner orbits.

\begin{table}
\centering
\caption{Overview of identifications and basic parameters of VW~LMi. RV is the radial velocity of the system's barycentre. The last column gives the reference: (1) - \citet{gaiadr2}, (2) - \citet{pribulla2008}, (3) - \citet{hipp2}, (4) - \citet{tycho}, (5) - \citet{mass}, (6) - \citet{pribulla2006}. The $(B-V)$ color index is given in the Tycho passbands, which are close to the Johnson $B$ and $V$.}
\label{tab1}
\begin{tabular}{cccl}
\hline
\multicolumn{4}{c}{VW~LMi} \\
\hline
GSC             &                    & 2519-2347        &     \\
HD              &                    & 95660            &     \\
HIP             &                    & 54003            &     \\
TIC ID          &                    & 166647000        &     \\
$\mu_\alpha$	& [mas.yr$^{-1}$]    & 13.303(120)	    & (1) \\
$\mu_\delta$    & [mas.yr$^{-1}$]    & $-6.083$(166)	& (1) \\
RV              & [km~s$^{-1}$]      & $-$0.15(25)      & (2) \\
$\pi$           & [mas]              & 7.47(70)         & (3) \\
$\pi$           & [mas]              & 9.05(12)         & (1) \\ 
$V_{\rm max}$   & [mag]              & 8.0003(123)      & (1) \\
$(B-V)$         & [mag]              & 0.340(21)        & (4) \\
$(J-K)$         & [mag]              & 0.208(30)        & (5) \\
$T_{\rm eff}$   & [K]                &  6506            & (1) \\ 
sp. type        &                    & F3-5V            & (6) \\ 
\hline
\end{tabular}
\end{table}

\section{Perturbations of orbits in multiple systems}
\label{perturbations}

Third or multiple bodies orbiting a binary star cause perturbations of its orbit. From the time-scale and amplitude point of view, we can consider the short-period, long-period and secular (or apse-node) perturbations. If $P_{12}$ is the orbital period of the inner binary and $P_3$ is the period of the outer orbit in a triple system, the periods of the short, the long and the secular perturbations are $P_{12}$, $P_3$, and $P^2_3/P_{12}$, respectively. The resulting element perturbations have amplitudes $(P_{12}/P_3)^2$, $P_{12}/P_3$, and unity, respectively \citep[see][]{hilditch}.

The three-body perturbation problem can be also applied to quadruple stellar systems. In the case of a 2+2 hierarchy we can approximate one binary system as a mass-point distant perturber and study its effects on the orbital evolution of the other binary and vice versa.

VW~LMi is the tightest quadruple system ever discovered with 2+2 hierarchy \citep{tokovinin}. The ratio of the outer (mutual) orbital period and inner orbital periods is an important parameter indicating the stability and the expected amplitude of perturbations. In the case of VW~LMi, the ratio of outer period $P_{1234}$ and period of the DNEB, $P_{34}$, is equal to $44.5$, which means that the system is stable \citep[see e.g.,][]{Toko18}. The time scale of secular orbital changes in the orbit of the DNEB, $P_{1234}^2/P_{34} \sim 43$ years, is now comparable with the observations span. In addition to the dynamically-excited apsidal motion, the expected perturbations in the system include orbital inclination changes estimated by \citet{vokrouhlicky} at $\Delta i_{12}$ = 0.2-0.4$\degr$ for the CEB and $\Delta i_{34}$ = 3-4$\degr$ for DNEB.

On the other hand, the RV and timing-variability analysis of \citet{pribulla2008} showed that except the dynamically-induced apsidal motion in the DNEB, no perturbations could be detected in 10 years of observations. The analysis showed that the inclinations of the inner and outer orbits are similar. Together with obvious stability of the orbits this indicates that the system is close to being co-planar so we cannot expect high-amplitude Kozai-Lidov perturbations, which occur if mutual inclination of the inner and outer orbits is larger than 39.2$\degr$ \citep[see][]{kozai,kiseleva,kim14}.

\section{New observations of VW~LMi}
\label{newobs}

Ground-based observations of VW~LMi are possible from mid latitudes of the Northern hemisphere from early November till late May. Relatively short observing season combined with 355-days outer orbital period means that full coverage of the outer orbit with ground-based observations requires about 15-18 years. Moreover, almost one-year outer period is expected to mimic parallactic motion and to complicate determination of the distance.\footnote{Because the ecliptic latitude of WW~LMi is $b$ = 22.3$\degr$ the shape of the parallactic ellipse would be the same as that of the photocentric motion for the outer orbit for inclination $90\degr- b=67.7\degr$.}

\subsection{Spectroscopic observations}
The spectroscopic observations were obtained from November 2016 till January 2020. Only spectra with the signal-to-noise ratio (SNR) higher than 20 in the yellow-green range were used.

Medium and high-dispersion spectroscopy of VW~LMi was obtained with three spectrographs. The observations at Star\'a Lesn\'a (SL) Observatory were performed at the G1 pavilion with a 60cm, f/12.5 Zeiss Cassegrain telescope equipped with a fiber-fed \'echelle spectrograph eShel \citep{2011IAUS..272..282T,ech}. The spectrograph has a 4150-7600 \AA~ (24 \'echelle orders) spectral range and a maximum resolving power of about $R$ = 11,000. The ThAr calibration unit provides about 100 m~s$^{-1}$ RV system stability. An Atik 460EX CCD camera, which has a 2749$\times$2199 array chip, 4.54 $\mu$m square pixels, read-out noise of 5.1~e$^-$ and gain 0.26e$^-$/ADU, was used as the detector. The observations were also performed with a 1.3m, f/8.36 Nasmyth-Cassegrain telescope equipped with a fiber-fed \'echelle spectrograph at the Skalnat\'e Pleso (SP) Observatory. Its layout follows the MUSICOS design \citep{1992A&A...259..711B}. The spectra were recorded by an Andor iKon-L DZ936N-BV CCD camera with a 2048$\times$2048 array, 13.5$\mu$m square pixels, 2.9e$^-$ read-out noise and gain close to unity. The spectral range of the instrument is 4250-7375 \AA~ (56 \'echelle orders) with the maximum resolution of $R$ = 38,000. Additional spectra were obtained at Th\"uringer Landessternwarte Tautenburg (TLS) with the Alfred Jensch 2m telescope and coud\'e \'chelle spectrograph. These spectra cover 4510-7610 \AA~ in 51 orders. A 2 arcsec slit was used for all observations giving R = 31,500. Because of the short orbital period of the contact pair the exposure times were limited to 900 seconds to prevent orbital-motion smearing of spectral lines.

The raw spectroscopic data were reduced using IRAF package tasks, Linux shell scripts and FORTRAN programs as described in \citet{ech}. In the first step, master dark frames were produced. In the second step, the photometric calibration of the frames was done using dark and flat-field frames. Bad pixels were cleaned using a bad-pixel mask, cosmic hits were removed using the program of \citet{2004PASP..116..148P}. Order positions were defined by fitting 6th order Chebyshev polynomials to tungsten-lamp and blue LED spectra. In the following step, scattered light was modeled and subtracted. Aperture spectra were then extracted for both object and ThAr lamp and then the resulting 2D spectra were dispersion solved. Finally, 2D spectra were combined to 1D spectra.

\subsection{Photometric observations}
The photometric observations were obtained from November 2017 till April 2019. A typical point-to-point scatter of the photometry was 0.01 mag.

\begin{figure}
\includegraphics[width=\columnwidth]{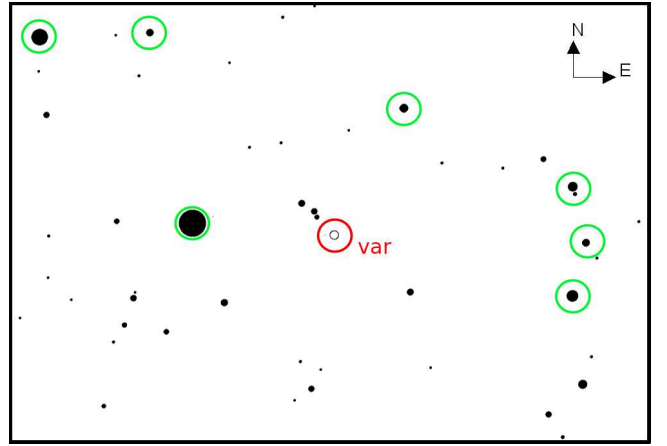}
\caption{Comparison stars used in aperture photometry and VW~LMi (denoted as "var"). Field of view dimensions are 25$\times$17 arcmin.}
\label{fig:1}
\end{figure}

The data were obtained at the G1 pavilion of the SL Observatory with a 18cm f/10 auxiliary Maksutov-Cassegrain telescope attached to the Zeiss 60cm Cassegrain used to obtain the \'echelle spectroscopy. An SBIG ST10 XME CCD camera (2184$\times$1472 array, 6.8$\mu$m square pixels, 11e$^-$ read-out noise and gain of 2.30e$^-$/ADU) and the Johnson-Cousins $BVRI$ filter set was used. The observations were typically taken in two filters during a given night. The field of view (FoV) of the CCD camera is 28.5$\times$18.9'. The CCD frames were photometrically reduced under the IRAF environment. First, master dark and master flat-field frames were produced, then bad pixels were cleaned and the frames were photometrically calibrated. Prior to aperture photometry all frames were astrometrically solved to define the pixel-to-WCS\footnote{World Coordinate System} transformation. Ensemble aperture photometry was performed with respect to 7 stars close to VW~LMi (see Fig.~\ref{fig:1}).

\section{Analysis of the spectra}
\label{spectra}

The analysis of the new spectroscopy of VW~LMi was focused on the determination of RVs of its components. No attempt to disentangle spectra was made.

The RVs were determined using the broadening-function (hereafter BF) technique developed by \citet{rucinski92}. The BFs have been extracted from the 4900-5510 \AA~ spectral range (free of hydrogen Balmer lines and telluric lines). The velocity step in the extracted BFs was set according to the spectral resolution. For eShel at G1 the step of $\Delta$RV = 5.8 km~s$^{-1}$ was used, for the MUSICOS at SP and Coud\'e \'echelle at TLS it was $\Delta$RV = 3 km~s$^{-1}$. The BF technique is particularly appropriate for VW~LMi because the spectral type of all four components is close. The BFs for eShel and MUSICOS spectra were extracted using HD128167 (F4V, RV = +0.2 km~s$^{-1}$) as the template while for the TLS data HD22879 (G0V, RV = +114.2 km~s$^{-1}$) was used. For spectra taken outside spectroscopic conjunctions BFs contain all four components: two rapidly rotating components of the CEB and two slowly rotating components of the DNEB. 

Assuming synchronous rotation, aligned rotational axes ($i_\star = i_{34}$) and $R_3 = R_4 = 1 R_\odot$ (corresponding to the estimated G2V spectral type) we get $v \sin{i_\star} = 6.2$ km~s$^{-1}$ for either of the components. The width of the observed profiles is consistent with the instrumental profile of the MUSICOS. Thus $v \sin{i_\star} < 8$ km~s$^{-1}$, which supports the synchronous rotations.

The light ratio of the fourth and third component defined by the Gaussian profile areas is $L_4/L_3 = 0.894\pm0.004$. The relative light contribution of the DNEB is $(L_3+L_4)/(L_1+L_2) = 0.484\pm0.008$ determined from the BFs obtained within $\pm$0.05 in phase from the CEB maxima. The light-ratio estimates do not take into account possible differences of the spectral types and their effect on the metallic lines strength \citep[see][]{rucinski13}.

\begin{figure}
\centering
\includegraphics[width=\linewidth]{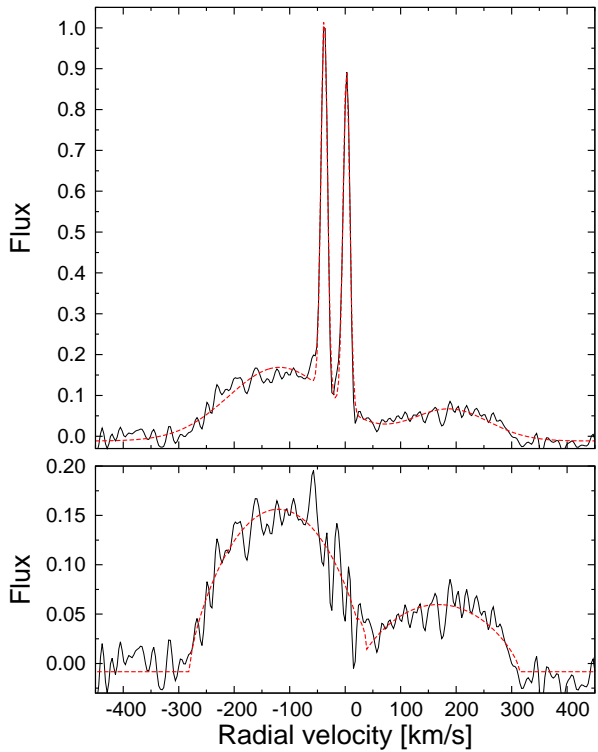}
\caption{Broadening function of VW~LMi extracted from a MUSICOS spectrum obtained on January 27, 2017 (top panel) and its best four-component Gaussian function model (red dashed line). The residual broadening function (bottom panel) and its double-rotational profile fit (red dashed line). The orbital phase of the CEB is $\phi_{12}$ = 0.190, while the DNEB is observed at $\phi_{34}$ = 0.388 measured form the periastron passage. The broadening function was normalized to unity.}
\label{fig2}
\end{figure}

The extracted BFs were first fitted by a multi-component Gaussian-function model, which was appropriate to define the RVS of the DNEB components. In the next step, the DNEB's Gaussian models have been subtracted from the original BFs. The resulting cleaned BFs were modeled by a limb-darkened rotational profiles more appropriate for rapid rotators (see Fig.~\ref{fig2}). The center positions of the rotational profiles defined
RVs of the CEB components. 


One of the drawbacks of the BF formalism is that the extracted BFs must be smoothed. The sigma of the smoothing Gaussian kernel should correspond to the spectral resolution but often more smoothing is required. Thus fits to the extracted and smoothed BFs do not provide reliable RV error estimates. To estimate the measurement errors and to arrive at useful errors of the spectroscopic elements, we first set errors as 1/SNR of the spectrum. The errors were then scaled so the orbital solution for a given component resulted in reduced $\chi^2$ = 1.

In the case of the TLS spectroscopy only two spectra provided RVs of the contact pair while all 8 spectra enabled us to measure RVs of the DNEB components. Although the TLS data are of higher SNR than the MUSICOS spectra, they were not used in the orbital solution of the system.

The precision of the RV measurements is determined by the resolution of the spectrograph, the SNR, the observed spectral range, the density of spectral lines, and the projected rotational velocity of the observed object ($v\sin i_\star$). Because of the higher spectral resolution while similar SNR, the RVs derived from MUSICOS at SP are expected to be of higher precision. All the new RVs are given in Tables~\ref{tabA1}, \ref{tabA2}, and \ref{tabA3}.

\section{Analysis of the photometry}
\label{lc-changes}

The photometry of VW~LMi was focused on (i) obtaining timing information, and (ii) search for secular changes of the inclination angle of the CEB orbit.

\begin{figure}
\centering
\includegraphics[width=\linewidth]{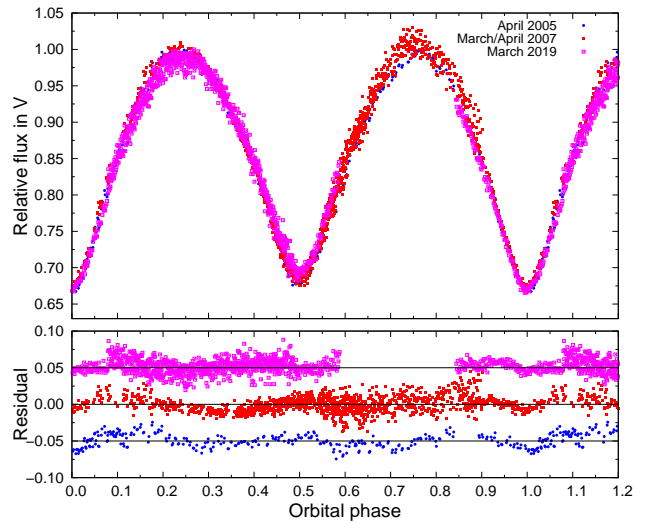}
\caption{Comparison of three $V$-band light curves of VW~LMi (top). The light curves were arbitrarily shifted so that the level of the maximum at phase 0.25 matches. The residuals from the best model fits obtained with Roche are plotted in the bottom panel.}
\label{fig-lc}
\end{figure}

The timing information was obtained by the same technique as in \citet{pribulla2008} fitting a passband-specific template LC to individual nightly LCs. The resulting minima times are always related to the preceding primary even if the secondary minimum was observed. In total, our photometry resulted in 29 new timings (see Table~\ref{tabA4}).

To investigate the possible precession of the CEB orbit, we modelled multi-colour observations from different epochs. The LC modelling was performed using the code Roche \citep{2004ASPC..318..117P}. Focusing on the inclination angle of the CEB, $i_{12}$, and trying to avoid parameter correlations, we adopted the parameters from the solution of \citet{djurasevic} for all LCs. Only the time of the primary minimum, temperature of the secondary component, and the inclination angle were adjusted.

Finally, three datasets were modelled: (i) our old $BV$ photoelectric data from April 2005 \citep{pribulla2008}, (ii) CCD $BVR$ photometry of \citet{djurasevic} from March/April 2007, and (iii) our new CCD $BV$ observations from March 2019. The $V$ passband LCs from these datasets are shown in Fig.~\ref{fig-lc}.

The LCs shows variability with changes at the level of 0.01-0.02. This variability is, very probably, caused by surface activity of the components \citep[see][]{djurasevic}. Neglecting the surface inhomogeneities in the LC modelling gives the following inclination angles: $i_{12}$ = 76.8.5$\pm0.2\degr$ ($BV$ photometry from April 2005), $i_{12}$ = 77.8$\pm0.2\degr$ \citep[$BVR_c$ photometry of][from March/April 2007]{djurasevic}, and $i_{12}$ = 76.1.5$\pm0.3\degr$ (new $BV$ data from March 2019). The most recent light curve shows, however, a shallower secondary minimum, while the primary minimum remains of a similar depth as in the 2005 and 2007 data. This means that the LC variations are  activity-dominated. This is supported by the significant differences in the temperature of the secondary component resulting from variations in the relative depth of the minima.
The surface inhomogeneities are well visible in the systematic residuals from the fits (Fig.~\ref{fig-lc}).

In the view of the expected inclination-angle variations of about 0.2-0.4$\degr$ (Vokrouhlick\'y, 2016, private communication), the surface activity effects dominate and prevent us from detecting the real inclination-angle changes.

VW~LMi (TIC ID 166647000) will be observed by the TESS \citep{TESS} in sector 22 from February 19 to March 17, 2020 with a 2 minute cadence\footnote{see https://tess.mit.edu/observations/sector-22/}. While the observing run is probably too short to disclose the perturbations of the CEB, the observations will enable to improve geometric elements and shed more light on the surface activity of the CEB. It is possible that a modulation of the CEB LC by rotation of the components of the DNEB will also be detected.

\section{Determination of the orbital parameters}
\label{orbit}

\begin{figure}
\includegraphics[width=\columnwidth]{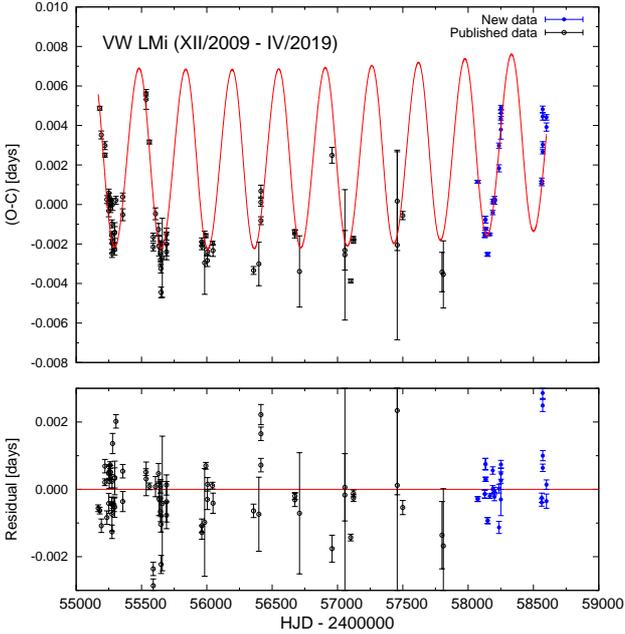}
\caption{Recent part of the ($O-C$) diagram and the best fit of the data corresponding to the simultaneous modeling with the four radial-velocity datasets (top) and the corresponding residuals (bottom)}
\label{fig4}
\end{figure}

From the set of RVs, orbital parameters of the quadruple system were determined via least-square fit of the RV equation to observational data. To obtain more precise values of the orbital parameters (see Table~\ref{tab2}), we fitted the observed times of photometric minima together with the observed RVs. In the case of 2+2 hierarchy of VW~LMi the RVs of the components can be expressed as \citep[see][]{pribulla2008}:

\begin{equation}
\label{eq4}
\begin{aligned}
RV_l &= V_0 + K_{12}[e_{1234}\cos\omega_{1234}+\cos(\nu_{1234}+\omega_{1234})]\\ 
&+(-1)^{l+1} K_l[e_{12}\cos\omega_{12}+\cos(\nu_{12}+\omega_{12})],
\end{aligned}
\end{equation}

\begin{equation}
\label{eq5}
\begin{aligned}
RV_l &= V_0 - K_{34}[e_{1234}\cos\omega_{1234}+\cos(\nu_{1234}+\omega_{1234})]\\ 
&+(-1)^{l+1} K_l[e_{34}\cos\omega_{34}+\cos(\nu_{34}+\omega_{34})].
\end{aligned}
\end{equation}

where $l$=1,2 in the first equation (CEB) and $l$=3,4 in the second equation (DNEB). The CEB's elements are denoted with index "12", while the DNEB's elements are denoted with index "34". The elements of the mutual orbit of the quadruple system are denoted with index "1234". The semi-amplitudes of the RV changes are component-specific and denoted in a different way: $K_l$ are semi-amplitudes of individual components in the inner orbits, $K_{12}$ and $K_{34}$ are semi-amplitudes of the RV changes of the mass centers of the binaries. $V_0$ is the systemic velocity of the whole quadruple system.

The observed minima times of the CEB were modelled using the equation:

\begin{equation}
\label{eq7}
T_{\rm min} = T_{12} + P_{12}E + Q E^2 +\Updelta T_{12},
\end{equation}

here $T_{12}$ is the time of the primary minimum at zero epoch ($E$ = 0), $P_{12}$ the period, $Q$ the coefficient of the quadratic ephemeris, $E$ the epoch and $\Updelta T_{12}$ is the delay caused by the LiTE:

\begin{equation}
\label{eq8}
\begin{aligned}
\Updelta T_{12} = &\frac{K_{12}P_{1234}(1-e_{1234})^{3/2}}{2\pi c} \times \\
 &\frac{\sin(\nu_{1234}+\omega_{1234})}{1+e_{1234}\cos\nu_{1234}}.
\end{aligned}
\end{equation}

Orbital parameters were determined via simultaneous fit of all RV datasets and minima times through optimization of the merit function ($\chi^2$) via least-squares method of multi-dimensional function minimization by \citet{nelder}. We took into account the apsidal motion effects in the detached pair with $d \omega/dt$ = const. The resulting orbital period is now anomalistic, i.e. from periastron to periastron. No other perturbations of orbits were assumed in the modelling. Continuous period increase visible in the timing data of the CEB was used in the RV data modelling. The times of the RV observations were corrected for the LiTE in the outer orbit as in \citet{pribulla2008}.

Spectra from SL and SP have significantly different quality. Thus, we decided to fit the data from these observatories separately. For both separate fits the same photometric minima times were used to better define the outer 355-days orbit. In the case of the minima timings, all available literature and new data since HJD 2\,455\,168 (December 2009) till HJD 2\,458\,599 (April 2019) were used. All minima used in the modelling are listed in Table~\ref{tabA3}.

While the RV scatter for the contact-binary star is similar for eShel and MUSICOS, for the DNEB the higher spectral resolution of MUSICOS makes the profiles of its components narrower resulting in a higher RV precision. The data uncertainties are not affected by the RV stability of the instruments which is about 100 m~s$^{-1}$.

\begin{figure}
\centering
  \includegraphics[width=8.2cm]{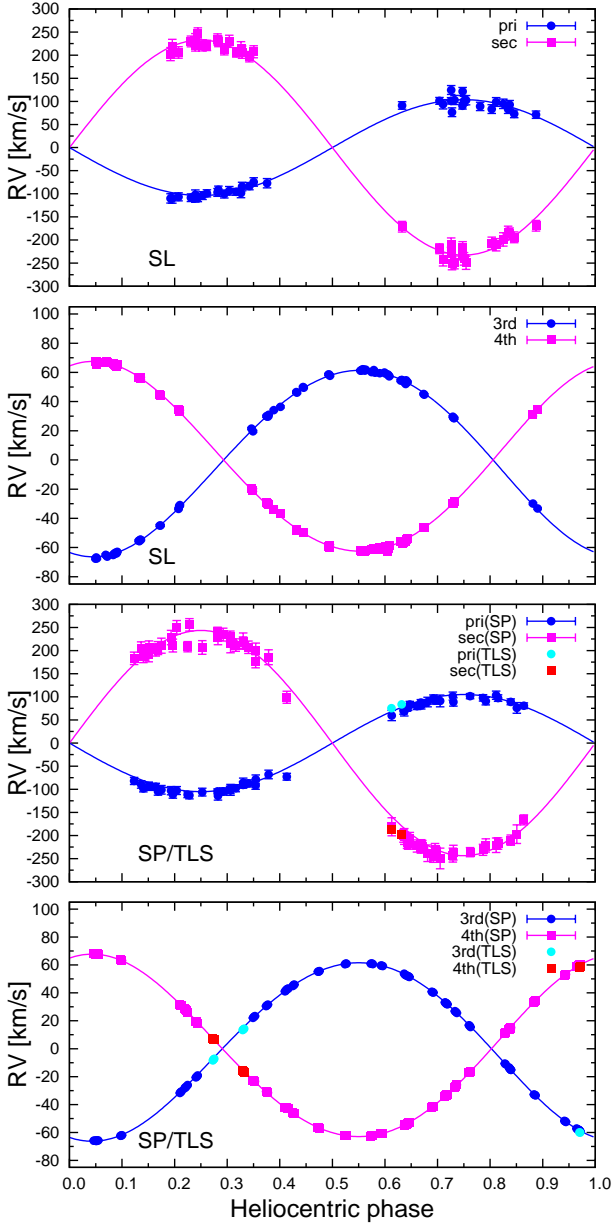}
   \caption{Best fits of VW LMi's radial velocities from 2016-2018 eShel, 2017-2020 MUSICOS, and 2016-2017 TLS data assuming no apsidal motion in the DNEB, but including the continuous period change of the CEB. The RVs were corrected for the motion in the outer 355-days orbit. The phases were computed from the primary minimum (CEB) and the periastron passage (DNEB). The TLS data were not used in the modelling and are plotted without error estimates.}
\label{fig5}
\end{figure}

The complete set of the new orbital parameters of VW~LMi is presented in Table \ref{tab2} and the corresponding best fits are shown in Figures~\ref{fig4} and \ref{fig5}. The 1998-2008 data presented in \citet{pribulla2008} were re-analysed the same way as the new observations. In this case, the minima timings from December 1997 till May 2008 were used.

New spectroscopy was obtained with smaller telescopes (60cm and 1.3m) than the data analysed in the paper of \citet{pribulla2008}, which were obtained with the 1.88m telescope at the DDO. The most problematic are RV measurements of the secondary component of the CEB. Its profile is very poorly defined, especially in the case of the eShel data, which will not be considered in the further analysis.

Minimum masses obtained from the inner and outer orbits indicate that the orbits could be close to being co-planar. From the MUSICOS data analysis we have $(m_1 + m_2) \sin^3i_{12}$ = 2.096$\pm$0.036 M$_\odot$ (inner orbit), and $(m_1 + m_2) \sin^3i_{1234}$ = 1.91$\pm$0.10 M$_\odot$ (outer orbit). In the case of the DNEB we have $(m_3 + m_4) \sin^3i_{34}$ = 1.7720$\pm$0.0023 M$_\odot$ (inner orbit), and $(m_3 + m_4) \sin^3i_{1234}$ = 1.61$\pm$0.14 M$_\odot$ (outer orbit). From the above results we have $\sin i_{12}/\sin i_{1234}$ = 1.0315$\pm$0.0189 (CEB) and $\sin i_{34}/\sin i_{1234}$ = 1.0325$\pm$0.0299 (DNEB). To determine the mutual inclination angle we would have to know the difference of the longitudes of the ascending nodes (see Eqn.~\ref{mutual}).

\begin{table}\scriptsize
\centering
\caption{Orbital parameters from the global fit to the radial velocities and the minima times. Heliocentric Julian Dates are given without 2\,450\,000. Average standard deviations for single observations, $\sigma_{1234}$ and $\sigma_{MIN}$ are also listed.}
\label{tab2}
\begin{tabular}{c|cc|cc|cc|c}
\hline \hline
\textbf{Parameter}       &  \textbf{Unit}  & \textbf{DDO}        & \textbf{eShel}        & \textbf{MUSICOS}      \\ 
Time range of RVs        &                 & 1998-2008           & 2016-2018             & 2017-2020             \\
\hline
$P_{12}$                 &   [days]        & 0.47755108(4)       & 0.4775503(3)          &  0.47755104(22)       \\
$Q$                      &   [days]        & 1.75(8) 10$^{-10}$  & 2.31(17) 10$^{-10}$   & 1.91(13) 10$^{-11}$   \\
$T_{12}$                 &   [HJD]         & 2\,500.14945(16)    & 2\,500.1529(12)       & 2\,500.1499(9)       \\
$K_1$                    &   [km/s]        & 105.5(11)           & 103.7(16)             & 105.0(13)             \\
$K_2$                    &   [km/s]        & 249.8(14)           & 232.8(24)             & 243.6(22)             \\
$m_{1} \sin^3 i_{12}$    &  [M$_\odot$]    & 1.560(23)           & 1.305(34)             & 1.465(33)             \\
$m_{2} \sin^3 i_{12}$    &  [M$_\odot$]    & 0.659(12)           & 0.581(17)             & 0.631(15)             \\
$a_1 \sin i_{12}$        &  [R$_\odot$]    & 0.996(10)           & 0.979(15)             & 0.991(12)             \\ 
$a_2 \sin i_{12}$        &  [R$_\odot$]    & 2.358(13)           & 2.197(23)             & 2.299(21)             \\ \hline
$P_{34}$                 &   [days]        & 7.93251(25)         & 7.9335(6)             & 7.93276(5)            \\
$T_{34}$                 &   [HJD]         & 2\,218.96(8)        & 2\,218.1(4)           & 2\,218.55(4)          \\
$e_{34}$                 &                 & 0.0351(3)           & 0.0402(17)            & 0.0385(4)             \\
$\omega_{34}$            &   [rad]         & 2.23(9)             & 1.10(28)              & 1.569(22)             \\
$d\omega_{34}/dt$        &   [rad/yr]      & 0.069(9)            & 0.113(18)             & 0.0800(15)            \\
$K_3$                    &   [km/s]        & 63.96(18)           & 63.97(12)             & 63.94(4)              \\
$K_4$                    &   [km/s]        & 65.54(21)           & 65.08(14)             & 65.35(4)              \\
$m_{3} \sin^3 i_{34}$    &  [M$_\odot$]    & 0.902(6)            & 0.888(6)              & 0.8957(16)            \\
$m_{4} \sin^3 i_{34}$    &  [M$_\odot$]    & 0.880(6)            & 0.873(6)              & 0.8763(16)            \\
$a_3 \sin i_{34}$        &  [R$_\odot$]    & 10.022(28)          & 10.021(19)            & 10.015(6)             \\
$a_4 \sin i_{34}$        &  [R$_\odot$]    & 10.27(3)            & 10.195(22)            & 10.236(6)             \\ \hline
$P_{1234}$               &   [days]        & 355.27(14)          & 355.98(18)            & 354.96(9)             \\
$T_{1234}$               &   [HJD]         & 3\,048(5)           & 3\,013(20)            & 3\,028(3)             \\
$K_{12}$                 &   [km/s]        & 20.8(6)             & 21.2(11)              & 21.0(9)               \\
$K_{34}$                 &   [km/s]        & 23.6(3)             & 23.26(21)             & 24.9(4)               \\
$e_{1234}$               &                 & 0.097(9)            & 0.027(6)              & 0.087(7)              \\
$\omega_{1234}$          &   [rad]         & 2.23(9)             & 2.1(4)                & 2.14(3)               \\
$m_{12} \sin^3 i_{1234}$ &  [M$_\odot$]    & 1.69(8)             & 1.69(9)               & 1.91(10)              \\
$m_{34} \sin^3 i_{1234}$ &  [M$_\odot$]    & 1.49(9)             & 1.54(16)              & 1.61(14)              \\
$a_{12} \sin i_{1234}$   &  [R$_\odot$]    & 145(4)              & 149(8)                & 147(6)                \\
$a_{34} \sin i_{1234}$   &  [R$_\odot$]    & 164.9(21)           & 163.6(15)             & 174.0(28)             \\ \hline
$V_0$                    &   [km/s]        & $-$0.62(22)         & $-$2.35(29)           & 0.4(3)                \\
\hline
$\sigma_{RV1}$           &   [km/s]        & 11.5                & 8.9                   &  7.0                  \\
$\sigma_{RV2}$           &   [km/s]        & 14.0                & 12.5                  & 13.2                  \\
$\sigma_{RV3}$           &   [km/s]        & 1.95                & 0.67                  & 0.24                  \\
$\sigma_{RV4}$           &   [km/s]        & 1.35                & 0.85                  & 0.29                  \\
$\sigma_{MIN}$           &   [days]        & 0.00084             & 0.00092               & 0.00092               \\
\hline
$\chi_{\nu} ^{2}(RV1)$   &                 & 1.02                & 1.00                  & 0.98                  \\
$\chi_{\nu} ^{2}(RV2)$   &                 & 1.01                & 0.97                  & 1.00                  \\
$\chi_{\nu} ^{2}(RV3)$   &                 & 1.01                & 1.01                  & 0.97                  \\
$\chi_{\nu} ^{2}(RV4)$   &                 & 1.02                & 0.98                  & 0.98                  \\
$\chi_{\nu} ^{2}(MIN)$   &                 & 1.03                & 1.04                  & 1.03                  \\
\hline
\end{tabular}
\end{table}

If we adopt the inclination angle of the CEB as 78.1$\pm0.3\degr$ \citep{djurasevic}, the true mass of the contact pair is 2.237$\pm$0.039 M$_\odot$, where the error is dominated by the RV modelling. From $K_{34}/K_{12} = 1.186\pm0.054$, the mass of the the DNEB is then $m_3 + m_4$ = 1.89$\pm$0.09 M$_\odot$. If the minimum mass of the DNEB is $(m_3 + m_4) \sin^3i_{34}$ = 1.7720$\pm$0.0023 M$_\odot$ then $i_{34} = 78.2^{+5.5}_{-3.9}$ degrees. The error of the inclination angle is dominated by the error of $K_{34}/K_{12}$. Combining true masses of the binaries $m_1 + m_2 + m_3 + m_4 = 4.13\pm$0.10 M$_\odot$ and the minimum total mass from the outer-orbit solution, $(m_1 + m_2 + m_3 + m_4) \sin^3 i_{1234} = 3.52\pm$0.17 M$_\odot$ we get $i_{1234} = 71.5^{+3.3}_{-2.9}$ degrees. Unfortunately, the multi-dataset solution does not set a strong constraint on the inclination angle of the DNEB orbit.

The apsidal-motion rate in the detached pair found from the simultaneous modelling corresponds to apsidal-motion cycle of $U$ = 78.6$\pm$1.6 years for the MUSICOS data but 91$\pm$11 years for the older data \citep{pribulla2008}\footnote{Originally estimated at 80 years in their paper}. The new MUSICOS data, however, resulted in a significantly higher precision of the RVs for the third and fourth component.

It is interesting to note, that except the marked apsidal motion most orbital elements of either inner and outer orbit remained constant within their errors during the whole interval of the observations. Most notably, semi-amplitudes of the RV changes of the components of the DNEB $K_3, K_4$ and the orbital parameters of the outer orbit.

\section{Numerical model of the system}
\label{nbody}

To ascertain the perturbations of the orbits of VW~LMi and to compare the predictions with observations, the numerical integration of the system was performed. We used an object-oriented gravitational modelling simulator developed by one of the authors (EP) in Python. The simulator is based on the Cowell's method of orbit integration and has a basic fourth-order Runge-Kutta (RK) integrator. We integrated the equation of motion for $N$ gravitationally interacting bodies, where the total acceleration affecting the $i$-th body is given by: 

\begin{equation}
\label{eq10}
\ddot{\vec{r_i}}  = -G \sum_{j=1, i \neq j}^{j=N}  \frac{m_j (\vec{r_i}-\vec{r_j})}{\vert \vec{r_i} - \vec{r_j} \vert^{3}},
\end{equation}

where $G$ is the gravitational constant, $m_j$ and $r_j$ is the mass and position vector of the $j$-th component. 

The integrations were performed assuming point mass approximation of the components of either of the binaries neglecting high-order terms of the gravitational potential caused by the tidal deformation, as well as the mass transfer between the components or stellar winds which also affect secular evolution of orbits.

Prior the numerical integration the orbital parameters of VW~LMi obtained from 2017-2020 MUSICOS and 2009-2019 timing data were transformed to the Cartesian position and velocity vectors at a given initial epoch using equations of celestial mechanics. These vectors, together with the masses of all four components were used as initial conditions in the numerical integration. The following masses and inclinations of orbits with respect to the plane of the sky were adopted: $m_1$ = 1.563 M$_\odot$, $m_2$ = 0.674 M$_\odot$, $m_3$ = 0.954 M$_\odot$, $m_4$ = 0.936 M$_\odot$, $i_{12}$ = 78.1$\degr$, $i_{34}$ = 68.9$\degr$ and $i_{1234}$ = 64.1$\degr$. Besides our simulator, we used C/Python simulator ReboundX \citep{tamayo2016reboundx} to perform additional long-term numerical integrations in order to study secular orbital changes in the orbit of the DNEB, which are present on timescales of decades. ReboundX uses a fifteenth-order adaptive step-size RK integrator \textit{ias15} \citep{rein2014ias15}. We integrated the quadruple system for up to 10\,000 outer orbital periods (9700 years).

Although it is probable that all orbits have the same sense of revolution, all possible cases were investigated: for the fixed inclination of the outer orbit there are two possible inclinations of the CEB $i_{12}$ and 180$\degr - i_{12}$ and two possible inclinations of the DNEB with $i_{34}$ and 180$\degr - i_{34}$. Therefore, for the fixed $i_{1234}$ there are 4 different cases.

The amplitude of the perturbations strongly depends on the mutual inclination of the orbits. This can be unambiguously found only when RV measurements are combined with positional observations. In this case we can determine longitude of ascending node, $\Omega$. For two orbits with inclination angles $i_1$, and $i_2$, and corresponding longitudes of ascending nodes, $\Omega_1$ and $\Omega_2$, the mutual inclination, $j_{1-2}$ can be determined as \citep[see e.g.][]{muter2006}:

\begin{equation}
\label{mutual}
\cos j_{1-2} = \cos i_1 \cos i_2 + \sin i_1 \sin i_2 \cos (\Omega_1 - \Omega_2).
\end{equation}

Because the mutual inclination (and perturbations) depend on the difference of the ascending nodes of the orbits, $\Omega_{1234}$ was set rather arbitrarily to zero. To investigate all possibilities, the numerical integrations were done for $\Omega_{34}$ = 0$\degr$ up to 180$\degr$ with a 30$\degr$ step and the same for $\Omega_{12}$ = 0$\degr$ up to 180$\degr$ with a 30$\degr$ step. This was performed for four possible inclinations of the inner binaries.

Cartesian positions and velocities obtained from the each set of integrations were then transformed back to osculation Keplerian elements. The comparison of the resulting time evolution of the osculation elements with the observations implies, that $\Omega_{34} \in \langle -30\degr, +30\degr \rangle$ and all orbital planes must be close to being co-planar. In non-coplanar cases the integrations predict large changes of orbital parameters which are not observed.

\begin{figure} 
    \centering
    \includegraphics[width=0.9\linewidth]{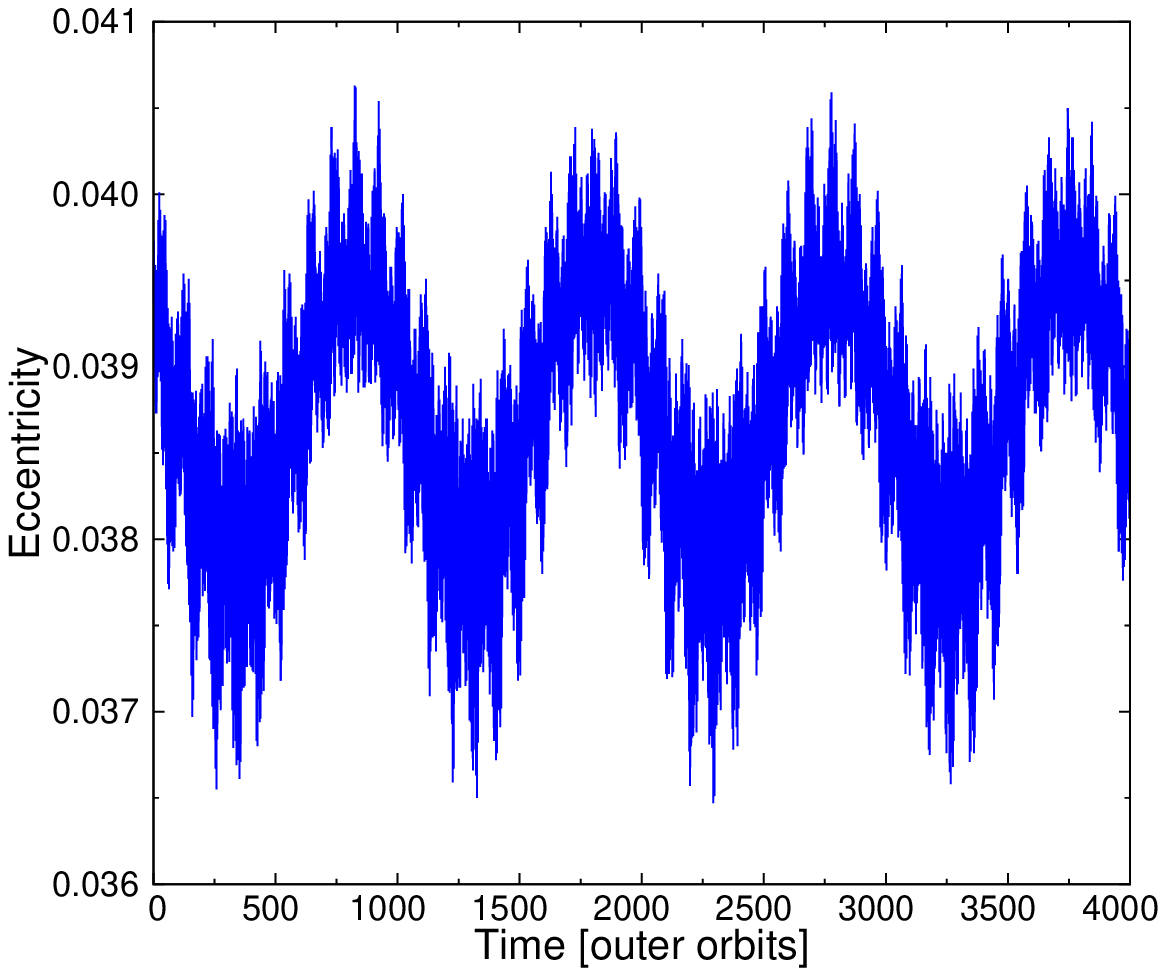} 
    \includegraphics[width=0.9\linewidth]{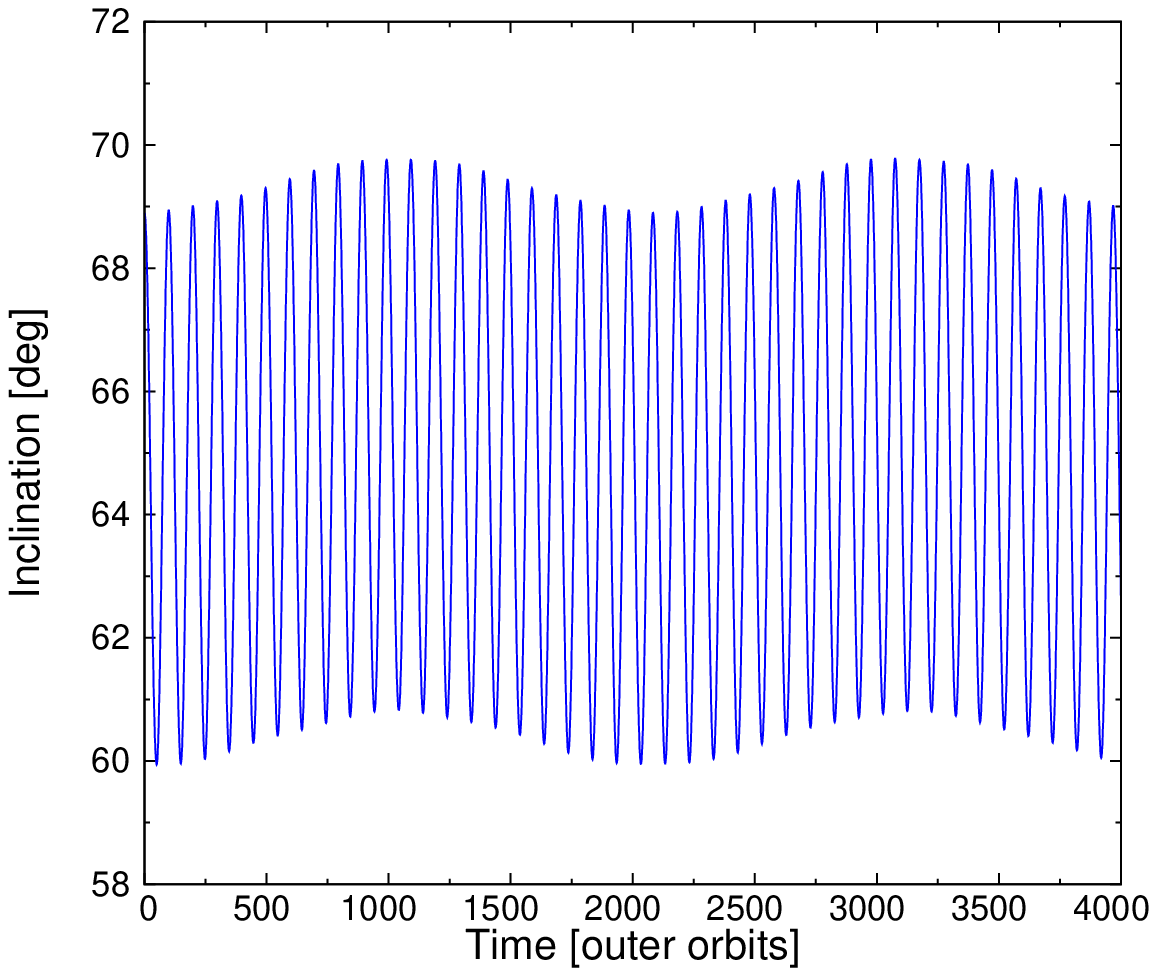} 
    \includegraphics[width=0.9\linewidth]{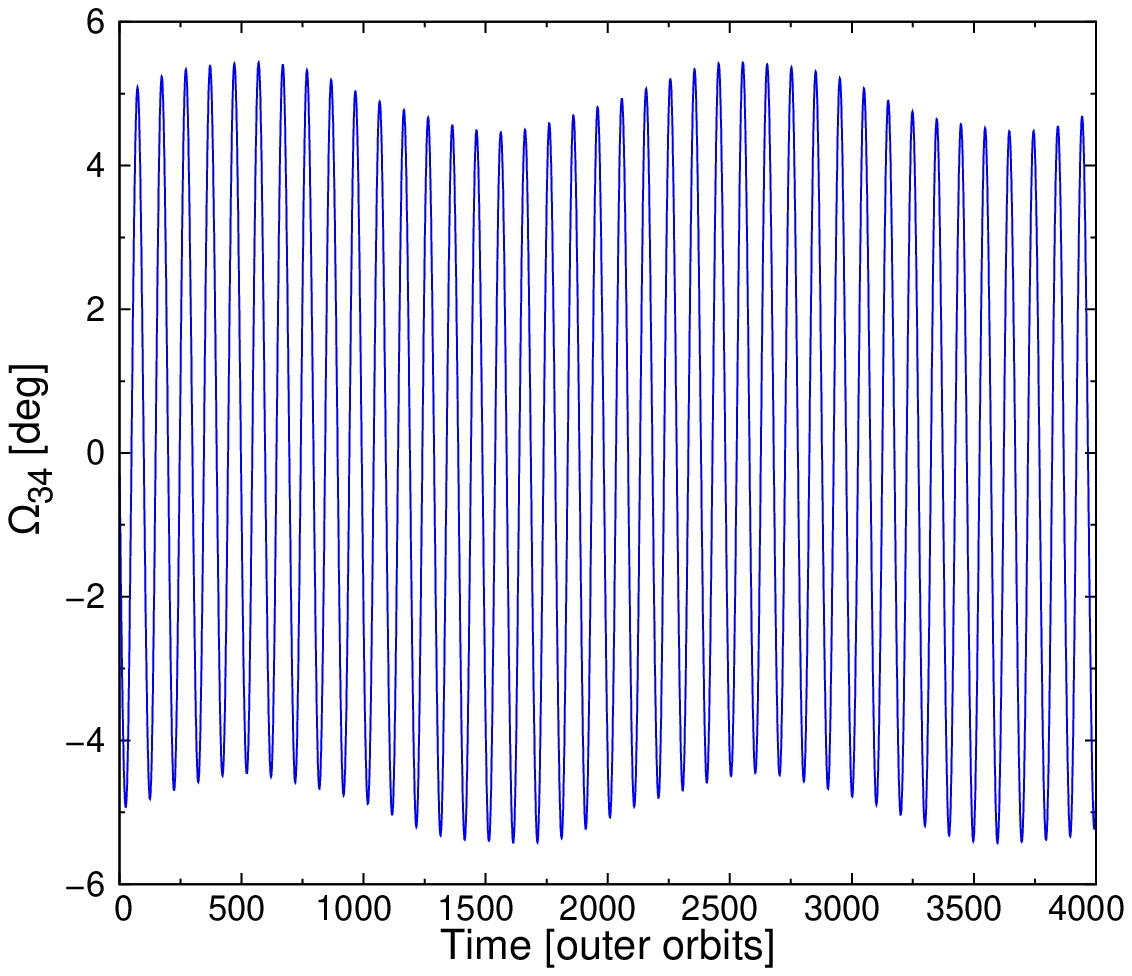}
  \caption{Long-term evolution of the orbital elements of VW~LMi's DNEB's orbit for initial value of $\Omega_{34} = 0$. Time units are in the orbital periods of the outer orbit (355 days).}
  \label{evolution34} 
\end{figure}

\begin{figure} 
    \centering
    \includegraphics[width=\linewidth]{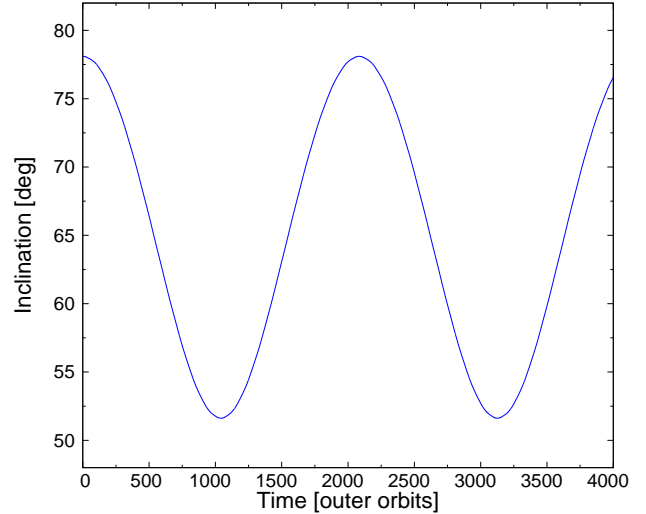} \caption{Long-term evolution of inclination angle of the CEB as a result of a point-mass gravitational simulation. Time units are in the orbital periods of the outer orbit (355 days).}
  \label{evolution12} 
\end{figure}


Evolution of the orbital elements of the DNEB is shown in Figure \ref{evolution34} for the initial value of $\Omega_{34} = 0\degr$.

\section{Evolution of the orbits}
\label{evol}

Changes of orbital elements are expected in the orbits of inner binary systems (as they perturb each others orbits) as well as in the outer orbit. Because the outer orbit carries the vast majority of the angular momentum, its perturbations are expected to have small amplitudes and long time scales.

\subsection{Changes in the DNEB's orbit}

The most pronounced orbital perturbation seen in the VW~LMi system is the fast apsidal motion of the DNEB orbit. Analysis of the new MUSICOS data (see Table~\ref{tab2}) gives the apsidal-motion cycle of $U$ = 78.6$\pm$1.6 years. The classical apsidal-motion cycle can be estimated using the equations 4.20 and 4.21 of \citet{hilditch} as $U \sim 308\,000$ years. The relativistic apsidal motion cycle estimated using the equation 4.31 of \citet{hilditch} is $U \sim 33\,400$ years. Thus the observed rate is dynamically induced with negligible relativistic and tidal contributions.

The apsidal-motion rate, caused by perturbations in a triple system, depends primarily on the inner-to-outer orbital period and mass ratio. Using the equations 7 and 8 of \citet{2007MNRAS.379..370B} which assume the orbital coplanarity, and component masses obtained from the MUSICOS data, we get the apsidal-motion cycle of 96.2 years. Using the higher mass of the contact pair corresponding to older 1998-2008 DDO data we get $U$ = 93.5 years. It is interesting to note, that due to the high mass of the perturber (CEB) and relatively large $P_{34}/P_{1234}$, the second term in the series for $d\omega/dt$ still contributes by about 10\%. Because the periods are well defined, getting at the observed apsidal-motion rate we would have to increase the contact-binary mass by about 20\%.

The mutual inclination of the orbits also affects the apsidal- motion rate. The equation 2.7 of \citet{borkovits2016} shows that if the orbits were mutually inclined by 20$\degr$, the apsidal-motion rate could increase by almost 12\%. The apsidal-motion rate found by our numerical integrations is the largest in the prograde case when the ascending-node difference, $(\Omega_{34} - \Omega_{1234})$ = 0 and $j_{34-1234}$ = 4.8$\degr$. In this case $U$ = 95.5 years. For the retrograde possibility of $j_{34-1234}$ = 57$\degr$ we get $U$ = 117 years.

Other orbital elements of the DNEB in VW~LMi seem to be constant when the results from the 1998-2008 and 2017-2020 observations are compared. A very important observational fact is that the semi-amplitudes of the RV changes remained constant within the errors. Comparing the projected semi-major axes from the 1998-2008 and 2017-2020 data allows the maximum inclination angle change of $\Delta i_{34} \leq $ 0.9$\degr$. The precession period \citep[see Eq. 8 of][]{pribulla2008} is about 120 years. The amplitude of the inclination-angle changes depends on the mutual inclination of the outer and inner orbit. Clearly, a more precise RVs of the DNEB are needed. Together with only a marginal increase of the eccentricity, $e_{34}$, it is clear that $(\Omega_{34} - \Omega_{1234}) \sim $ 0 (see Fig.~\ref{evolution34}) and orbits must be close to co-planar.

\subsection{Changes in the CEB's orbit}

The orbital parameters of the CEB in VW~LMi (see Table \ref{tab2}) do not show any significant changes between the 1998-2008 DDO and 2017-2020 MUSICOS data. Unfortunately, the spectroscopic elements are negatively influenced by the presence of the DNEB in the BFs decreasing the RV accuracy. Moreover, the profiles of the contact pair are poorly defined in the BFs, which probably causes the apparent change of the semi-amplitudes $K_1, K_2$. 

A much more sensitive indicator of the inclination-angle changes is the depth of the eclipses \citep[see the case of the Kepler eccentric binaries in][]{borkovits2015}. As analysed and discussed in Section~\ref{lc-changes}, the LC shape is influenced by the surface activity \citep[see also][]{dumitrescu2}, which virtually prevents us to detect the inclination-angle changes caused by the gravitational perturbations. The lack of significant perturbations also means that the outer and the contact-binary orbits are close to being coplanar excluding the retrograde possibility with $i_{12}$ = 180 - 78$\degr$.

Our numerical integrations show that the maximum inclination-angle changes are below 1$\degr$ during the time range of the observations. The integrations predict a cyclic inclination-angle change with the period of about 2000 years and the amplitude of about $\pm 13\degr$ (depending on the mutual inclination of the inner and outer orbit) caused by the precession of the orbit (Fig.~\ref{evolution12}).

\subsection{Changes in the outer orbit}

In the case of the mutual orbit, the timescale of the perturbations is much larger than the time interval of the available observations. The orbital elements obtained from the 1998-2008 RV data (1998-2008 timings) modelling are mostly within 1-sigma errors of the elements obtained from the recent 2017-2020 RV data (2009-2019 timings). 

\subsection{Expected secular perturbations}
\label{secular}

Long-term integrations of VW~LMi indicate that we should be able to detect secular changes of parameters (see Fig.~\ref{evolution34}). The inclination-angle amplitudes depend on the mutual angles between the inner and the outer orbit. In the case of the CEB, the apparent inclination angle should vary between 78$\degr$ (now) and a minimal value of about 52$\degr$ as a result of the CEB orbit precession with about 2000-year cycle. The inclination-angle of the DNEB should change much faster due to the smaller period ratio. The numerical integration predicts a combination of about 100-years cycle and about 2000-years modulation. The time evolution of the apparent inclination angle of the DNEB orbit shows another, about 6000 year cycle. Similar cycles are visible in the longitude of the ascending node of DNEB. The apsidal-motion rate is practically constant due to near co-planarity of the DNEB and outer orbit.


\section{Distance to the system}

VW~LMi is composed of main-sequence stars. While the trigonometric parallax might be affected by the photocentric motion, the observed maximum brightness and flux ratios of the system provide a 
possibility to separately determine brightness of the components and to estimate the distance to the system.

The photometric solution of the $BVR$ photometry of \citet{djurasevic} showed that the third light, $l_3/(l_1 + l_2 + l_3)$, increases towards red being 0.24, 0.26 and 0.28 in the $B$, $V$ and $R$ passband, respectively. Using the third light and the TYCHO magnitudes (Table~\ref{tab1}) we get the apparent maximum brightness of the contact pair as 8.73 mag in the $B$, and 8.39 mag in the $V$ passband. 

The absolute magnitude of the contact pair can be estimated from the period-colour-magnitude relation of \citet{Rucinski97} which for $P$ = 0.47755 days, and $(B-V)_0 = 0.34$ mag (assuming no extinction towards VW~LMi) gives $M_V$ = 2.57 mag and corresponding distance about 146 pc or parallax $\pi$ = 6.85 mas. Using a more recent period-magnitude calibration of \citet{Rucinski17} based on the Gaia DR1 TGAS (Tycho-Gaia astrometric) solutions for 297 contact-binary systems with $0.275<P<0.575$ days (relation \#4) gives $M_V$ =  3.04 mag with expected uncertainty of about 0.06 mag. Using apparent brightness of the contact pair, $V$ = 8.39 mag, the distance to the system is 117$\pm$4 pc, corresponding to the parallax of $\pi$ = 8.54$\pm$0.28 mas.

Light-curve modelling in Section~\ref{lc-changes} gives even larger distance to VW~LMi. Assuming spectroscopic elements obtained form the MUSICOS data ($K_1 + K_2$ = 355 km~s$^{-1}$, the mass ratio $q = m_2/m_1 = 0.4223$), and adopting $T_1$ = 6790 K, results in $M_V$ = 2.48 mag corresponding to distance of about 152 pc ($\pi$ = 6.57 mas). Getting at the Gaia DR2 parallax, $\pi$ = 9.05 mas, the temperature of the contact pair would have to be significantly lower, about 5900 K, which however corresponds to much redder colour than observed.

Both calibrations and the system modelling show that the Gaia parallax is affected by the multiplicity of the system where the orbital motion mimics a larger parallax.


2MASS near-infrared photometry of VW~LMi was obtained at HJD2\,450\,894.8450 at phase 0.953 close to the primary minimum. The observed $(J-K)$ = 0.208(30) mag corresponds to F2V with about one subclass uncertainty \citep[see][]{Allen2000} excluding much later spectral type required by the Gaia parallax.

A reasonable observed colours and third light can be obtained combining an F0V spectral type CEB with two G2V stars. Using theoretical brightness and colour of main-sequence stars \citep[see][p388]{Allen2000} and our model absolute visual magnitude of the CEB, $M_V$ = 2.48 mag, gives third light in the $V$ passband $l_3/(l_1 + l_2 + l_3)$ = 0.241 consistent with determination of \citet{djurasevic}.

\section{Conclusions}

VW~LMi is the tightest quadruple stellar system with the 2+2 hierarchy yet discovered. Both of the system's inner binaries act as perturbers on each other, causing secular orbital changes on time scales as short as decades. The analysis of the system is complicated by the surface activity of the CEB and a heavy blending of the component's profiles.

The main results of this study are:

\begin{itemize}

\item Determination of new orbital elements of both inner binaries and masses of all four components.

\item The inclination of the inner and outer orbits found from the analysis of the spectroscopic and timing data do not exclude their coplanarity.

\item Reliable determination of the apsidal-motion period of the DNEB, $U$ = 78.6$\pm$1.6 years, and finding that the classical tidal and relativistic contributions are negligible. The observed apsidal-motion rate is more consistent with the prograde orientation of the orbits.

\item Finding that semi-amplitudes of the RV changes, $K_3$ and $K_4$, of the DNEB remained constant within the errors for the 1998-2008 DDO and the 2017-2020 MUSICOS data. This means that the inclination angle of the DNEB orbit is practically constant ($\Delta i_{34} \leq 0.9\degr$) and indicates that the outer 355-days orbit is close to be co-planar to the DNEB orbit. 

\item Detection of a small but statistically significant increase of eccentricity of the DNEB orbit, $e_{34}$ (see Table~\ref{tab2}). The eccentricity of the DNEB's orbit increased by 0.0034(5) between the 1998-2008 DDO data and the 2017-2020 MUSICOS data. This is consistent with the marginal detection in Table~6 of \citet{pribulla2008}.

\item Finding that the outer orbit is stable within the observation errors indicating an absence of an outer perturber to the quadruple system.

\item Comparison of the numerical integrations of the system with observed parameter changes of the system. The observed apsidal-motion rate, and constancy of other parameters during the available range of the observations support close coplanarity of the orbits within about 5$\degr$.

\item Finding that perturbations of the contact-binary orbit are smaller than the detection limits. This is primarily caused by difficulties in measuring RVs of the wide profiles of the contact pair and its surface activity which prevents accurate measurement of the minima depth.

\item Estimate of the spectral types of the components as F2V for the contact pair and G2V for the components of the non-eclipsing binary.

\item Finding that the Gaia DR2 parallax is significantly affected by the photocenter motion. The system is probably much further than the available analysis of the Gaia astrometry (DR2) indicates.

\end{itemize}

The most intriguing question that remains to be answered is the formation of such a tight system. The existence of the CEB could be a consequence of the Kozai-Lidov cycles in an originally misaligned system. Normally, the transport of the angular momentum to additional component could bring the present contact pair to the regime where tidal friction combined with magnetic braking could further shrink the orbit. The problem is, that the third component (DNEB) remains close to the CEB orbiting it in less than one year.

This tightness of the outer orbit could have been caused by an extra and yet unknown companion of an originally quintuple system,  which was expelled during a close encounter event and "stole" most of the angular momentum. If this encounter occurred between one of the binaries it could change the orientation of its angular momentum vector. The observations available now, however, indicate that angular momentum vectors of the inner pairs are within a few degrees from the angular momentum of the outer orbit.

Better characterization of VW LMi calls for long-term monitoring to cover the whole 355-d orbital cycle. Obtaining times of minima with higher accuracy and covering as large part of the yearly observing season is important to better define the total mass of the CEB. Further, high-resolution spectroscopy is necessary to get more insight on the perturbations of the DNEB orbit. The understanding of the system would greatly benefit from visual orbit obtained by means of long-baseline interferometry. Unfortunately, the system is rather faint for Northern facilities such as the CHARA interferometer.

The photocenter motion of VW~LMi, mimicking parallactic motion, will complicate the astrometric solution of the Gaia observations. A reliable solution might be possible only towards the end of the mission when the difference between yearly parallactic component and the 355-day orbital motion will accumulate.

\section*{Acknowledgements}

The authors thank to V. Kollár and P. Sivanič for the technical assistance. This work was supported by the VEGA grant of the Slovak Academy of Sciences No. 2/0031/18 and the VEGA grant of Comenius University  No. 1/0911/17. This work has  also been  supported  by  the grant  of  the  Slovak  Research and Development Agency number APVV-15-0458. T.\,B. acknowledges the financial support of the Hungarian National Research, Development and Innovation Office -- NKFIH Grant KH-130372. The authors thank to an anonymous referee for his/her constructive comments.
\bibliographystyle{mnras}
\bibliography{biblio} 


\appendix
\onecolumn

\section{New radial velocities and minima times of VW~LMi}
\label{appendixA}

\begin{table}
\small
\centering
\caption{New radial velocities obtained from spectra taken with eShel spectrograph at the Star\'a Lesn\'a Observatory. The estimated signal-to-noise ratio (SNR) at 5500 \AA~ is given.}
\label{tabA1}
\begin{tabular}{rrrrrrrrrrrr} \hline \hline
\multicolumn{1}{c}{HJD}  & RV1     & RV2     & RV3    & RV4    & SNR  & \multicolumn{1}{c}{HJD}  & RV1    & RV2     & RV3    & RV4    & SNR  \\
2\,450\,000+ & {[}km.s$^{-1}${]}   & {[}km.s$^{-1}${]}& {[}km.s$^{-1}${]} & {[}km.s$^{-1}${]}& & 2\,450\,000+     & {[}km.s$^{-1}${]} & {[}km.s$^{-1}${]}& {[}km.s$^{-1}${]}&{[}km.s$^{-1}${]}&\\ \hline
7753.61306 & -         & -         &     40.45 &  $-$18.71 & 57.9 &  8133.54619 &     -      &    -    &   55.55 & $-$53.66 & 49.0 \\
7753.62351 & -         & -         &     39.62 &  $-$18.98 & 58.6 &  8133.55664 &     -      &    -    &   54.14 & $-$53.55 & 46.2 \\
7753.63396 &     78.57 & $-$200.87 &     39.43 &  $-$17.40 & 51.0 &  8188.38640 &  $-$94.78  &  217.67 &   43.53 & $-$79.75 & 52.8 \\
7774.59330 &    117.22 & $-$216.42 &     33.22 &  $-$25.55 & 47.9 &  8188.39764 &  $-$96.66  &  241.53 &   43.52 & $-$80.38 & 55.6 \\
7774.60375 &    113.99 & $-$223.23 &     32.99 &  $-$26.29 & 52.5 &  8188.40816 &  $-$92.06  &  232.81 &   43.90 & $-$79.58 & 54.7 \\
7774.61536 & -         & -         &     32.88 &  $-$27.20 & 46.8 &  8188.42183 &  $-$85.24  &  245.90 &   44.42 & $-$79.82 & 54.1 \\
7774.62583 & -         & -         &     34.21 &  $-$27.47 & 51.3 &  8188.43250 &  $-$81.78  &  241.86 &   43.80 & $-$79.50 & 49.4 \\
7780.59550 & $-$105.65 &    207.49 &  $-$54.53 &     57.62 & 53.6 &  8188.44316 &  $-$86.72  &  226.93 &   44.01 & $-$80.14 & 45.8 \\
7780.60644 & $-$100.95 &    200.70 &  $-$54.81 &     57.33 & 63.4 &  8188.45546 &     -      &    -    &   43.43 & $-$80.23 & 47.8 \\
7780.61828 &  $-$88.74 &    191.36 &  $-$53.68 &     57.39 & 58.1 &  8188.46675 &  $-$65.21  &  134.03 &   44.08 & $-$80.03 & 46.1 \\
7780.62932 & -         & -         &  $-$53.75 &     56.67 & 53.7 &  8188.55633 &     -      &   -     &   42.38 & $-$78.87 & 48.6 \\
7782.63349 & -         & -         &     34.22 &  $-$33.57 & 47.5 &  8188.58909 &     -      &   -     &   43.79 & $-$78.69 & 52.7 \\
7797.50532 &    101.81 & $-$218.11 &     -     &     -     & 61.6 &  8188.60135 &     -      &   -     &   42.30 & $-$79.47 & 51.6 \\
7797.51577 &    101.91 & $-$225.12 &     -     &     -     & 65.9 &  8188.61235 &     -      &   -     &   42.00 & $-$79.37 & 45.6 \\
7797.52622 &     91.24 & $-$232.47 &     -     &     -     & 66.3 &  8200.40752 &     -      &   -     &$-$85.92 &    45.90 & 52.2 \\
7800.48913 & -         & -         &     47.68 &  $-$61.63 & 54.6 &  8200.41800 &     -      &   -     &$-$86.40 &    46.64 & 47.1 \\
7800.49959 & -         & -         &     46.73 &  $-$61.34 & 61.9 &  8200.42847 &     -      &   -     &$-$87.14 &    45.70 & 48.0 \\
7800.51004 & -         & -         &     46.51 &  $-$60.51 & 52.9 &  8203.38315 &     -      &   -     &   28.35 & $-$71.56 & 55.4 \\
7814.45470 &$-$103.45  &    226.35 &     24.43 &  $-$48.50 & 45.1 &  8203.39362 &    106.38  &$-$155.91&   27.97 & $-$71.57 & 56.3 \\
7840.40455 & -         & -         &     25.31 &  $-$66.34 & 57.6 &  8217.36442 &     88.72  &$-$151.70&$-$57.54 &    10.44 & 56.6 \\
7840.41518 & -         & -         &     24.76 &  $-$66.37 & 52.5 &  8217.37487 &     -      &   -     &$-$55.80 &     9.58 & 53.5 \\
7843.38760 &  103.63   &$-$145.62  &  $-$88.27 &     46.68 & 50.7 &  8217.38539 &     -      &   -     &$-$54.93 &     9.15 & 47.3 \\
7843.39807 &   97.80   &$-$192.67  &  $-$87.49 &     46.63 & 48.1 &  8218.46518 &  $-$92.15  &  219.90 & $-$2.58 & $-$43.61 & 45.7 \\
7843.40868 &  109.10   &$-$184.27  &  $-$88.11 &     44.60 & 45.1 &  8218.48769 &  $-$84.25  &  250.04 & $-$4.48 & $-$44.70 & 45.5 \\
7844.35720 &  113.45   &$-$195.62  &  $-$65.82 &     24.29 & 48.5 &  8220.40013 &  $-$90.70  &  265.07 &   35.13 & $-$83.99 & 54.0 \\
7844.36816 &   97.65   &$-$168.22  &  $-$65.78 &     23.22 & 50.7 &  8236.33086 & -          & -       &   34.57 & $-$86.56 & 46.2 \\
7845.32873 &   87.92   &$-$179.27  &   -       &       -   & 56.0 &  8236.34671 & -          & -       &   34.05 & $-$84.93 & 49.4 \\
7854.34581 &   93.14   &$-$234.48  &     23.56 &  $-$70.80 & 48.6 &  8236.38217 &    112.39  &$-$223.18&   33.29 & $-$87.83 & 45.5 \\
7854.35665 &  117.74   &$-$224.23  &     23.17 &  $-$71.34 & 47.8 &  8236.39262 &    121.68  &$-$229.73&   32.75 & $-$86.29 & 46.5 \\
8133.48278 &$-$115.50  &   215.06  &     56.52 &  $-$53.47 & 52.0 &  8236.40307 &    121.41  &$-$230.72&   32.05 & $-$83.56 & 45.8 \\
8133.49323 &$-$105.42  &   214.80  &     56.94 &  $-$53.92 & 59.1 &  8251.34718 & -          & -       &   33.98 & $-$84.17 & 50.3 \\
8133.50368 & $-$97.77  &   223.30  &     56.93 &  $-$54.93 & 56.6 &  8251.35936 & -          & -       &   33.88 & $-$84.47 & 48.9 \\
8133.52527 & $-$90.65  &   200.14  &     56.10 &  $-$54.23 & 49.1 &  8251.37108 & -          & -       &   33.15 & $-$83.44 & 51.5 \\
8133.53574 & $-$80.72  &   200.99  &     55.74 &  $-$53.86 & 50.1 &  8262.36314 & -          & -       & $-$52.96&     7.91 & 46.7 \\
\hline
\end{tabular}
\end{table}                                                                       

\begin{table}\small
\centering
\caption{New radial velocities obtained from spectra taken with MUSICOS spectrograph at the Skalnat\'e Pleso Observatory. The estimated signal-to-noise ratio (SNR) at 5500 \AA~ is given.}
\label{tabA2}
\begin{tabular}{rrrrrrrrrrrr} \hline \hline
\multicolumn{1}{c}{HJD}  & RV1     & RV2     & RV3    & RV4    & SNR  & \multicolumn{1}{c}{HJD}  & RV1    & RV2     & RV3    & RV4    & SNR  \\
2\,450\,000+& {[}km.s$^{-1}${]} & {[}km.s$^{-1}${]}& {[}km.s$^{-1}${]} & {[}km.s$^{-1}${]}& & 2\,450\,000+  & {[}km.s$^{-1}${]} & {[}km.s$^{-1}${]}& {[}km.s$^{-1}${]}&{[}km.s$^{-1}${]}&\\ \hline
7781.47744 &  $-$98.99 &  186.24    & $-$18.62 &    21.49  & 49.1 &  8242.31006 & $-$61.49 &  203.78   &  $-$1.17 & $-$45.98 & 47.8 \\
7781.49066 & $-$105.42 &  204.72    & $-$17.96 &    21.21  & 45.5 &  8242.32354 & $-$73.30 &  228.07   &  $-$0.42 & $-$46.35 & 47.6 \\
7781.50415 & $-$112.20 &  210.19    & $-$17.25 &    20.30  & 45.8 &  8242.33798 &     -    & -         &     0.05 & $-$46.66 & 28.3 \\
7843.42113 &    92.28  & $-$180.83  & $-$84.33 &    49.54  & 31.8 &  8245.35449 &  -       & -         &     3.66 & $-$50.66 & 42.7 \\
7843.43494 & -         & -          & $-$84.04 &    49.66  & 34.8 &  8245.36799 &  -       & -         &     3.17 & $-$50.03 & 36.4 \\
7843.44809 & -         & -          & $-$83.87 &    49.46  & 24.9 &  8245.38314 &  -       & -         &     2.41 & $-$49.05 & 46.5 \\
8107.68792 & $-$121.70 &  198.33    &    43.42 & $-$17.78  & 57.3 &  8268.38151 & -        & -         &    33.21 & $-$75.28 & 35.3 \\
8107.70120 & $-$116.16 &  196.71    &    43.92 & $-$17.98  & 43.0 &  8268.39296 & -        & -         &    32.48 & $-$74.81 & 35.7 \\
8107.71543 & $-$124.29 &  218.78    &    44.54 & $-$18.71  & 39.6 &  8268.40653 & -        & -         &    32.20 & $-$74.74 & 39.3 \\
8155.57940 & -         & -          &    37.25 & $-$47.22  & 26.5 &  8268.41799 & -        & -         &    31.92 & $-$74.45 & 36.3 \\
8155.59305 & -         & -          &    37.67 & $-$47.78  & 27.1 &  8269.35827 & 119.60   & $-$217.95 &  $-$3.52 & $-$37.70 & 64.8 \\
8155.60620 & -         & -          &    37.94 & $-$48.43  & 30.7 &  8269.36993 & 115.75   & $-$212.90 &  $-$3.93 & $-$37.00 & 53.3 \\
8166.50074 &  $-$59.47 &  194.06    &     -    & -         & 39.4 &  8269.38193 & 121.53   & $-$202.94 &  $-$4.79 & $-$36.49 & 40.9 \\
8168.53354 &   79.28   & $-$191.72  & $-$75.60 &  58.28    & 46.6 &  8270.33879 & 115.20   & $-$199.64 & $-$52.89 &    13.29 & 52.1 \\
8168.54970 &   90.36   & $-$213.35  & $-$75.53 &  58.21    & 48.0 &  8270.35031 & 107.02   & $-$192.17 & $-$53.25 &    13.90 & 57.6 \\
8168.56363 &  102.00   & $-$230.64  & $-$75.71 &  57.95    & 44.7 &  8270.36220 &  98.56   & $-$148.35 & $-$53.64 &    14.55 & 59.3 \\
8168.57888 &   97.94   & $-$227.92  & $-$75.49 &  58.20    & 40.3 &  8530.50644 & -        & -         &    28.95 & $-$54.12 & 44.7 \\
8179.34841 &  $-$92.48 &  252.92    &    28.02 & $-$55.03  & 67.5 &  8530.51790 & -        & -         &    28.51 & $-$53.58 & 28.1 \\
8179.35987 &  $-$86.50 &  240.04    &    28.51 & $-$55.14  & 68.1 &  8530.52936 & -        & -         &    28.30 & $-$53.21 & 38.1 \\
8179.46652 & -         & -          &    31.80 & $-$59.01  & 41.8 &  8531.48010 &  88.82   & $-$206.29 &  -       &  -       & 53.3 \\
8179.47797 & -         & -          &    32.61 & $-$59.21  & 43.9 &  8531.49156 &  98.92   & $-$215.57 &  -       &  -       & 56.1 \\
8179.48944 & -         & -          &    32.78 & $-$59.66  & 47.0 &  8531.50302 & 107.91   & $-$233.88 &  -       &  -       & 49.5 \\
8186.44242 &  $-$75.86 &  218.45    &     -    &     -     & 45.3 &  8532.50568 & 102.09   & $-$210.03 & $-$64.15 &    39.93 & 45.0 \\
8186.45563 &  $-$80.84 &  215.24    &     -    &     -     & 48.0 &  8532.51713 & -        & -         & $-$64.52 &    40.35 & 31.2 \\
8186.46958 &  $-$89.19 &  241.75    &     -    &     -     & 41.3 &  8532.52860 & -        & -         & $-$65.01 &    40.66 & 40.3 \\
8212.36357 & -         & -          &    39.96 & $-$83.58  & 45.9 &  8555.40002 & -        & -         & $-$29.26 &  $-$7.93 & 46.6 \\
8212.37977 & -         & -          &    40.19 & $-$83.41  & 50.7 &  8555.41147 & -        & -         & $-$29.72 &  $-$7.35 & 49.4 \\
8212.39488 & -         & -          &    39.72 & $-$83.30  & 40.7 &  8562.42716 & -        & -         &    13.07 & $-$54.32 & 35.7 \\
8223.42798 & -         & -          & $-$80.47 &    37.16  & 41.5 &  8562.43862 & -        & -         &    12.74 & $-$53.71 & 31.3 \\
8223.44101 & -         & -          & $-$81.01 &    37.27  & 46.2 &  8562.45007 & -        & -         &    12.04 & $-$53.37 & 29.7 \\
8223.45566 &   95.88   & $-$195.32  & $-$81.63 &    37.76  & 49.9 &  8563.39948 & -        & -         & $-$34.03 &  $-$6.81 & 25.7 \\
8229.35421 & -         & -          &    10.00 & $-$56.79  & 42.4 &  8563.41094 & -        & -         & $-$34.99 &  $-$5.90 & 23.9 \\
8229.36716 & -         & -          &     9.60 & $-$55.92  & 37.8 &  8563.42240 & -        & -         & $-$35.35 &  $-$5.62 & 27.1 \\
8229.38048 & -         & -          &     8.94 & $-$55.53  & 44.4 &  8566.35932 &  107.11  & $-$219.99 & $-$52.51 &    11.21 & 35.6 \\
8230.30623 & -         & -          & $-$36.33 &  $-$9.02  & 50.5 &  8566.37078 &  109.51  & $-$231.18 & $-$51.87 &    10.42 & 28.6 \\
8230.32144 & -         & -          & $-$37.14 &  $-$8.28  & 71.4 &  8566.38225 &  119.96  & $-$223.91 & $-$51.38 &    10.10 & 42.1 \\
8230.33771 & -         & -          & $-$38.08 &  $-$7.46  & 54.8 &  8573.39483 &$-$53.64  &  118.28   & $-$84.24 &    41.92 & 49.3 \\
8231.32807 & -         & -          & $-$80.21 &    35.75  & 38.2 &  8573.40630 & -        & -         & $-$83.99 &    41.66 & 47.6 \\
8231.33952 & $-$73.00  &  212.12    & $-$80.62 &    36.04  & 40.2 &  8573.41776 & -        & -         & $-$83.80 &    41.38 & 52.1 \\
8231.35098 & $-$80.81  &  231.06    & $-$81.11 &    36.21  & 39.3 &  8574.36129 & -        & -         & $-$50.27 &     6.67 & 41.9 \\
8231.36495 & $-$82.29  &  270.54    & $-$81.23 &    36.50  & 42.7 &  8574.37275 & -        & -         & $-$49.62 &     6.14 & 39.0 \\
8231.37640 & $-$92.74  &  276.06    & $-$81.73 &    37.17  & 47.9 &  8574.38420 & -        & -         & $-$49.26 &     5.52 & 36.5 \\
8236.30959 & -         & -          &    36.44 & $-$83.92  & 28.1 &  8576.37185 &  103.00  & $-$185.57 &    32.97 & $-$78.58 & 53.7 \\
8236.32255 & -         & -          &    36.25 & $-$83.75  & 28.1 &  8576.38330 &  106.02  & $-$203.78 &    33.04 & $-$78.78 & 50.7 \\
8236.33553 & -         & -          &    35.95 & $-$83.97  & 29.7 &  8576.39475 &  111.73  & $-$211.33 &    33.35 & $-$79.11 & 46.5 \\
8241.32504 &  -        & -          & $-$49.31 &     2.87  & 21.6 &             &          &           &          &          &      \\
\hline
\end{tabular}
\end{table}

\begin{table}\small
\centering
\caption{New radial velocities obtained from spectra taken with Coud\'e \'echelle spectrograph of Th\"{u}ringer Landessternwarte at Tautenburg. The estimated signal-to-noise ratio (SNR) at 5500 \AA~ is given.}
\label{tabA3}
\begin{tabular}{rrrrrrrrrrrr} \hline \hline
\multicolumn{1}{c}{HJD} & RV1     & RV2     & RV3    & RV4    & SNR  & \multicolumn{1}{c}{HJD} & RV1    & RV2     & RV3    & RV4    & SNR  \\
2\,450\,000+ & {[}km.s$^{-1}${]} & {[}km.s$^{-1}${]}& {[}km.s$^{-1}${]} & {[}km.s$^{-1}${]}& & 2\,450\,000+  & {[}km.s$^{-1}${]} & {[}km.s$^{-1}${]}& {[}km.s$^{-1}${]}&{[}km.s$^{-1}${]}&\\ 
\hline
7739.67783 &  61.63   & $-$199.96   & $-$42.50 &  75.74  & 44.9 &  7742.55103 & -        & -        &  30.76 &  $-$0.48  & 75.2 \\
7739.68710 &  69.92   & $-$211.74   & $-$42.77 &  75.95  & 48.9 &  7797.59298 & -        & -        &   3.09 & $-$12.43  & 74.5 \\
7742.52941 & -        & -           &    29.82 &   0.68  & 49.3 &  7797.60371 & -        & -        &   2.65 & $-$11.80  & 74.0 \\
7742.54021 & -        & -           &    30.31 &   0.10  & 62.5 &  7797.61445 & -        & -        &   2.23 & $-$11.28  & 77.0 \\
\hline
\end{tabular}
\end{table}

\begin{table}\normalsize
\centering
\caption{Minima times used in the simultaneous analysis of the timing variability and radial-velocity observations.
Filter "Ir" is clear but IR blocking while "C" is clear or without a filter. The column Ref. gives the reference: (1) -  \citet{OEJV0137}, 
(2) -  \citet{IBVS5980}, (3) -  \citet{IBVS5974}, (4) -  \citet{IBVS6044}, (5) -  \citet{OEJV160}, (6) -  \citet{IBVS6007}, 
(7) -  \citet{IBVS6010}, (8) -  \citet{IBVS6033}, (9) -  \citet{OEJV165}, (10) -  \citet{IBVS6084}, (11) -  \citet{OEJV179},
(12) -  \citet{IBVS6149}, (13) -  \citet{IBVS6152}, (14) -  \citet{IBVS6196}, (15) -  \citet{IBVS6204}, (16) -  \citet{VSOLJ64}, 
(17) -  \citet{IBVS6244}, (18) -  this paper \label{tabA4}}
\begin{tabular}{llll|llll|llll}
\hline \hline
\multicolumn{1}{c}{HJD} & $\sigma$ & Fil. & Ref. & \multicolumn{1}{c}{HJD} & $\sigma$ & Fil. & Ref. & \multicolumn{1}{c}{HJD} & $\sigma$ & Fil. & Ref.  \\
 2\,450\,000+&         &    &      &   2\,450\,000+&          &   &      &   2\,450\,000+&          &    &         \\ 
\hline
 5168.71379 &  0.0001 &  R &  (1) &   5644.3502  &  0.0002  & V &  (4) &   7499.41168 &  0.00021 & R  & (15)    \\ 
 5180.6518  &  0.0001 &  C &  (1) &   5645.30454 &  0.0003  & V &  (5) &   7800.0293  &          & V  & (16)    \\ 
 5192.58931 &  0.0002 &  R &  (1) &   5645.30494 &  0.0005  & R &  (5) &   7810.2966  &  0.0017  & Ir & (17)    \\  
 5220.52522 &  0.0001 &  R &  (1) &   5648.40821 &  0.00023 & C &  (6) &   8073.67253 &  0.00007 & V  & (18)    \\ 
 5220.5257  &  0.0002 &  R &  (2) &   5649.36211 &  0.00027 & C &  (6) &   8127.63348 &  0.00012 & V  & (18)    \\ 
 5235.56593 &  0.0002 &  R &  (1) &   5659.3925  &  0.0020  & N &  (7) &   8133.36491 &  0.00017 & I  & (18)    \\ 
 5249.41444 &  0.0003 &  I &  (1) &   5691.38894 &  0.0004  & R &  (5) &   8134.79710 &  0.00006 & V  & (18)    \\ 
 5249.41534 &  0.0002 &  R &  (1) &   5691.38934 &  0.0006  & I &  (5) &   8149.60000 &  0.00009 & V  & (18)    \\ 
 5253.47405 &  0.0002 &  R &  (1) &   5691.38984 &  0.0003  & V &  (5) &   8073.67253 &  0.00007 & V  & (18)    \\ 
 5253.47435 &  0.0002 &  I &  (1) &   5961.44625 &  0.0002  & R &  (5) &   8127.63348 &  0.00012 & V  & (18)    \\ 
 5260.8762  &  0.0001 &  V &  (3) &   5961.44645 &  0.0002  & V &  (5) &   8133.36491 &  0.00017 & I  & (18)    \\ 
 5263.5024  &  0.0002 &  R &  (2) &   5983.4129  &  0.0016  & C &  (8) &   8134.79710 &  0.00006 & V  & (18)    \\ 
 5272.3358  &  0.0002 &  R &  (2) &   5992.48780 &  0.0001  & C &  (5) &   8149.60000 &  0.00009 & V  & (18)    \\ 
 5274.48388 &  0.0002 &  I &  (1) &   6004.4254  &  0.0003  & V &  (4) &   8166.79298 &  0.00007 & V  & (18)    \\ 
 5274.4844  &  0.0005 &  R &  (2) &   6005.38095 &  0.0002  & C &  (5) &   8188.76214 &  0.00011 & V  & (18)    \\ 
 5274.48538 &  0.0002 &  R &  (1) &   6043.34696 &  0.0001  & C &  (5) &   8188.76158 &  0.00011 & R  & (18)    \\ 
 5278.3068  &  0.0003 &  R &  (2) &   6047.4058  &  0.0003  & V &  (4) &   8201.65626 &  0.00011 & B  & (18)    \\ 
 5280.45389 &  0.0001 &  R &  (1) &   6357.57637 &  0.0002  & C &  (9) &   8201.65609 &  0.00012 & I  & (18)    \\ 
 5290.48229 &  0.0002 &  R &  (1) &   6398.4076  &  0.0011  & Ir& (10) &   8236.52041 &  0.00013 & B  & (18)    \\ 
 5294.3026  &  0.0003 &  R &  (2) &   6414.40786 &  0.0002  & B &  (5) &   8236.51926 &  0.00018 & I  & (18)    \\ 
 5295.49734 &  0.0003 &  I &  (1) &   6414.40879 &  0.0002  & V &  (5) &   8250.37085 &  0.00026 & B  & (18)    \\ 
 5295.49734 &  0.0005 &  R &  (1) &   6414.40936 &  0.0003  & R &  (5) &   8250.37029 &  0.00048 & I  & (18)    \\ 
 5304.33372 &  0.0002 &  R &  (1) &   6673.71922 &  0.0002  & B & (11) &   8251.32624 &  0.00017 & B  & (18)    \\ 
 5356.38642 &  0.0003 &  I &  (1) &   6673.71933 &  0.0001  & V & (11) &   8251.32651 &  0.00012 & I  & (18)    \\ 
 5356.38732 &  0.0002 &  R &  (1) &   6709.5339  &  0.0018  & C & (12) &   8563.16568 &  0.00011 & I  & (18)    \\ 
 5534.52032 &  0.0001 &  R &  (1) &   6958.58439 &  0.0004  & C & (11) &   8563.16582 &  0.00014 & B  & (18)    \\ 
 5535.71393 &  0.0005 &  R &  (1) &   7056.4780  &  0.0033  & Ir& (13) &   8570.33239 &  0.00018 & B  & (18)    \\ 
 5560.54460 &  0.0001 &  R &  (1) &   7056.7170  &  0.0010  & Ir& (13) &   8570.33276 &  0.00016 & V  & (18)    \\ 
 5589.43133 &  0.0002 &  I &  (1) &   7101.36679 &  0.0001  & C & (11) &   8571.28572 &  0.00011 & B  & (18)    \\ 
 5589.43183 &  0.0002 &  R &  (1) &   7122.38119 &  0.0001  & R & (11) &   8571.28608 &  0.00015 & V  & (18)    \\ 
 5607.3413  &  0.0003 &  R &  (2) &   7122.38133 &  0.0001  & V & (11) &   8599.46268 &  0.00021 & I  & (18)    \\ 
 5630.5019  &  0.0003 &  V &  (4) &   7457.3854  &  0.0048  & Ir& (14) &   8599.46317 &  0.00014 & B  & (18)    \\ 
 5633.36638 &  0.0005 &  R &  (5) &   7457.6264  &  0.0025  & Ir& (14) &              &          &    &         \\
\hline
\end{tabular}
\end{table}


\bsp	
\label{lastpage}
\end{document}